\documentclass[iop,apj]{emulateapj}
\usepackage{apjfonts}

\catcode`\@=11
\newcommand{\gapprox}{\mathrel{\mathpalette\@versim>}}
\newcommand{\lapprox}{\mathrel{\mathpalette\@versim<}}
\newcommand{\propapprox}{\mathrel{\mathpalette\@versim\propto}}
\newcommand{\@versim}[2]
  {\lower3.1truept\vbox{\baselineskip0pt\lineskip0.5truept
\ialign{$\m@th#1\hfil##\hfil$\crcr#2\crcr\sim\crcr}}}
\catcode`\@=12

\newcommand{\src}{G141.2+5.0}

\shorttitle{X-ray Source in Radio PWN G141.2+5.0}

\begin{document}

%%%%%%%%%%%%%%%%%%%%%%%%%%%%%
%%%%% Title of proposal %%%%% 
%%%%%%%%%%%%%%%%%%%%%%%%%%%%%

\title{A Compact X-ray Source in the Radio Pulsar-Wind Nebula G141.2+5.0}

\author{Stephen P. Reynolds\altaffilmark{1}
\&\ Kazimierz J. Borkowski\altaffilmark{1}
}

\altaffiltext{1}{Department of Physics, North Carolina State University, 
Raleigh, NC 27695-8202; reynolds@ncsu.edu} 

\begin{abstract}

We report the results of a 50 ks {\sl Chandra} observation of the
recently discovered radio object G141.2+5.0, presumed to be a
pulsar-wind nebula.  We find a moderately bright unresolved X-ray
source which we designate CXOU J033712.8 615302 coincident with the
central peak radio emission.  An absorbed power-law fit to the 241
counts describes the data well, with absorbing column $N_H = 6.7 (4.0,
9.7) \times 10^{21}$ cm$^{-2}$ and photon index $\Gamma = 1.8 (1.4,
2.2)$.  For a distance of 4 kpc, the unabsorbed luminosity between 0.5
and 8 keV is $ 1.7^{+0.4}_{-0.3} \times 10^{32}$ erg s$^{-1}$ (90\%
confidence intervals).  Both $L_X$ and $\Gamma$ are quite typical of
pulsars in PWNe.  No extended emission is seen; we estimate a
conservative $3 \sigma$ upper limit to the surface brightness of any
X-ray PWN near the point source to be $3 \times 10^{-17}$ erg
cm$^{-2}$ s$^{-1}$ arcsec$^{-2}$ between 0.5 and 8 keV, assuming the
same spectrum as the point source; for a nebula of diameter $13''$,
the flux limit is 6\% of the flux of the point source.  The steep
radio spectrum of the PWN ($\alpha \sim -0.7$), if continued to the
X-ray without a break, predicts $L_X\ \rm{(nebula)} \sim 1 \times
10^{33}$ erg s$^{-1}$, so additional spectral steepening between radio
and X-rays is required, as is true of all known PWNe.  The high
Galactic latitude gives a $z$-distance of 350 pc above the Galactic
plane, quite unusual for a Population I object.

\end{abstract}


\section{Introduction}

Pulsar-wind nebulae, the bubbles of relativistic particles and
magnetic field blown by pulsars, perform several important astrophysical
functions.  Most simply, they can serve as calorimeters for the energy
loss from pulsars, allowing the inference of unseen pulsars beamed
away from us.  But they also serve as laboratories for the study of
the behavior of highly relativistic flows at shock waves, so are
useful in the study of extragalactic jets and gamma-ray burst sources.
Pulsar-wind nebulae (PWNe) were originally defined by radio
properties: center-filled morphology, flat radio spectrum ($\alpha
\sim -0.3 - 0$, with $S_\nu \propto \nu^\alpha$) and (relatively) high
polarization -- properties originally found in only a handful of
objects (an early catalog, Weiler 1985, lists eight ``pure'' PWNe and another
five ``well-established'' PWNe inside radio shells).  However, the
launch of {\sl Chandra}, and to a lesser extent {\sl XMM-Newton,}
ushered in a new era in the discovery of pulsars and PWNe.  Finally,
the advent of TeV and GeV studies, primarily with H.E.S.S., MAGIC, and
VERITAS at TeV energies, and Fermi and Agile in GeV, has revealed a
trove of hard-spectrum, center-brightened sources, many of which turn
out to harbor pulsars, and (almost) all of which are therefore
presumed to be PWNe -- many at a much older age than previous objects,
old enough to have long outlived their natal supernova remnant (SNR)
shell.  Kargaltsev, Rangelov, \& Pavlov (2013) list 76 pulsars
containing X-ray and/or TeV PWNe.  See Gaensler \& Slane (2006) for a
general review of PWNe.

There are now enough PWNe known that general properties of the class
are fairly well understood -- or at least have become familiar.  The
initial pulsar wind seems to be ``dark,'' that is, cold in the fluid
frame and radiating inefficiently.  The wind is thermalized in some
fashion in a termination shock marking the inner edge of the observed
PWN (Rees \& Gunn 1974), where the characteristic bright synchrotron
emission (ranging from radio to X-ray wavelengths) appears.  Particles
are transported downstream into the nebula by some combination of
advection and diffusion, until the outer edge of the PWN where the
wind interacts either with the interior of a more-or-less ``normal''
shell SNR, or, if it is much older, with undisturbed ISM.  However,
none of those processes is well understood.

The wind leaving the pulsar is almost certainly dominated by Poynting
flux, but probably needs to be particle-dominated at the termination
shock, to allow there to be a shock at all (Kennel \& Coroniti 1984a).
The ratio of magnetic to particle energy flux, the
``magnetization'' $\sigma$, apparently must drop by orders of
magnitude in the dark zone. However, the spherical Kennel \& Coroniti (1984a)
hydrodynamic model does not
take account of the ``striped wind'' feature of an oblique rotating
neutron star, in which the azimuthal magnetic field beyond the light
cylinder changes direction with the pulsar's rotation.  Thus magnetic
reconnection is highly likely, and will affect the magnetization in
the dark zone.  Additionally, it is likely that the magnetization at
the termination shock is a function of latitude.  These effects can be
seen in recent 3D MHD simulations (Porth et al.~2013), in which
solutions exist with varying $\sigma$ and significantly higher average
values.

At the shock, the outgoing kinetic energy is somehow turned into
random particle energy. The spectrum of particles released into the
nebula, as inferred from the synchrotron spectral-energy distribution
(SED), appears initially quite flat ($N(E) \propto E^{-s}$ with $s
\equiv 1 - 2\alpha \sim 1 - 2$ for radio-emitting particles), but all
PWNe show steepening at higher photon energies.  The extent to which
this is due to intrinsic physics of the thermalization instead of
post-shock evolutionary effects is not known.  The initial particle
acceleration is probably not traditional diffusive shock acceleration,
since the shock is both relativistic and probably almost exactly
perpendicular, as the wind is expected to contain a very tightly wound
Parker spiral of magnetic field.  Particle acceleration by magnetic
reconnection is an attractive possibility, but quantitative
predictions are difficult.  In any case, the detailed nature of the
particle energization process remains obscure.

One major problem in understanding particle energization at PWN shocks
concerns the relation of radio emission to that at higher energies.
Kennel \& Coroniti (1984b) simply threw up their hands at the problem
of the Crab Nebula's radio emission, as it could not easily be
accommodated in their otherwise very successful scheme of an ideal
steady MHD outflow in spherical geometry.  Suggested solutions
typically invoke a totally separate particle population produced
through a separate process; e.g., Atoyan (1999), who proposes that the
radio-emitting electrons were injected early in the life of the nebula
and have simply aged since.  (This picture requires the Crab pulsar to
have been born with an initial rotation period of 3 -- 5 ms, a
striking assertion.)  Additionally, steepening of the spectrum between
radio and X-rays by an amount greater than the increase in power-law
index of 0.5 expected for synchrotron losses in homogeneous sources is
almost universally observed (e.g., data in Chevalier 2005).  Is this
an intrinsic property of the particle acceleration mechanism?  If so,
in the wind, at the termination shock, or elsewhere?  Advection models
can reproduce this steepening purely from evolutionary effects on an
initial straight power-law (Reynolds 2009) at the expense of invoking
ad-hoc gradients in source properties.  The comparison of X-ray and
radio properties is the most effective way to address these important
questions; ideally, X-ray observations can also reveal the powering
neutron star, even if it is not detectable in radio pulsations.

\section{\src}

\begin{figure}
\includegraphics[scale=0.4]{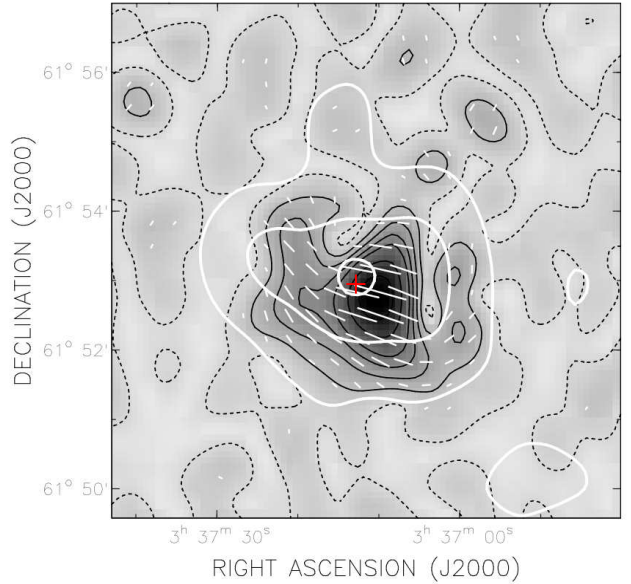}
\caption{\small {DRAO image in polarized intensity at 1420 MHz of
    G141.2+5.0 (Kothes et al.~2014).  The resolution is $56'' \times
    48''$.  Total-intensity contours are shown in white.  The bulge in
    the lowest contour is an unrelated point source The point-source
    position is indicated by the red cross (much larger than the
    positional uncertainty).}
\label{radio}}
\end{figure}

The discovery of additional PWNe has occurred in recent years
primarily at very high photon energies.  However, those objects have
turned out mostly to be far older and interacting directly with ISM,
introducing additional complications in modeling.  An alternative
approach is to start with objects of known radio properties, the
original defining characteristics of PWNe.  One such object, only
recently discovered, is G141.2+5.0, found in Canadian Galactic Plane
Survey (CGPS; Taylor et al.~2003) observations using the Dominion
Radio Astronomy Observatory (DRAO) at 1.4 GHz (Kothes et al.~2014,
hereafter K14).  This object, the first radio-discovered PWN in 17 years
(see Fig.~\ref{radio}), has a 1.4 GHz flux density of 0.14 Jy, and
center-brightened morphology with high radio polarization (15\%
integrated over the source, but reaching 40\% at peak).  It thus has
all the earmarks of a normal radio PWN -- except for the spectrum,
which is much steeper: $\alpha = -0.69 \pm 0.05$, more characteristic
of an extragalactic source (even a little too steep for typical shell
SNRs).  K14 review alternate possible interpretations of the source,
but the complete absence of obvious counterparts in infrared, optical,
or soft X-ray surveys rules out H II regions or nearby radio galaxies.
Cluster halos normally have too steep spectra ($\alpha \lapprox -1.0$)
while cluster relics have similarly steep spectra but also much less
spherical morphologies.  The discovery we report here of an X-ray
point source coincident with the intensity peak of \src\ essentially
confirms the PWN interpretation.

G141.2+5.0 shows the center-brightened morphology and substantial
(also center-brightened) linear polarization characteristic of the
class.  H I observations give a kinematic distance of $4 \pm 0.5$ kpc,
but also reveal a surrounding shell of H I with a radius of about
$6'$, expanding at 6 km s$^{-1}$ (K14).  Faraday rotation observations
imply a substantial amount of internal Faraday rotation, indicating
significant thermal ionized gas.  These latter properties, along with
the steep radio spectrum, make G141.2+5.0 a highly unusual PWN.

\begin{figure}
\centerline{\includegraphics[width=8truecm]{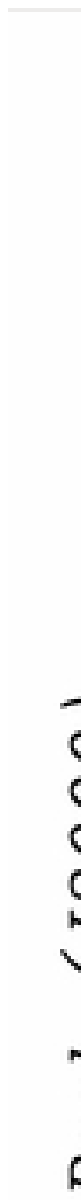}}
\caption{\small {DRAO image in total intensity at 1420 MHz of
    G76.9+1.0 (Landecker et al.~1993).  The resolution is $15'' \times
    14''$.  Contours near the center show the X-ray PWN (Arzoumanian et al.
~2011).}
\label{G76radio}}
\end{figure}

While the radio spectrum of G141.2+5.0 is anomalous for a PWN, K14
cite two other PWNe with steep radio spectra: G76.9+1.0 (Landecker et
al.~1993) and DA 495 (Kothes et al.~2008).  G76.9+1.0 shows a fairly
circular envelope of diameter about $7'$ enclosing two maxima in radio
(Fig.~\ref{G76radio}; Landecker et al.~1993), and a known pulsar
between them: with a period of 24 ms and a rotational energy-loss
$\dot E = 1.2 \times 10^{38}$ erg s$^{-1}$, it is the second most
energetic pulsar in the Galaxy (Arzoumanian et al.~2011).  (This
confirms that unprepossessing nebulae may contain unusual pulsars.)
It also contains a tiny ($16'' \times 10''$) X-ray nebula, shown as
the faint contours near the center in Fig.~\ref{G76radio}.  The small
size and low flux of its X-ray PWN may be typical for PWNe with steep
radio spectra; J2022+3842 has a very low efficiency $\eta$ of turning
spindown power into PWN X-ray luminosity: $L_X({\rm PWN}) \sim 6
\times 10^{32}$ erg s$^{-1}$ for a distance of 10 kpc (Arzoumanian et
al.~2011), and $\eta \sim 2 \times 10^{-5}$.  (Its magnetospheric
efficiency is about 10 times larger, that is, $L_X ({\rm PWN}) \sim
0.1 L_X ({\rm pulsar})$).  DA 495, with a radio extent of about $20'$,
also has a very small X-ray nebula (about $40''$), with a central
point source thought to be a pulsar, though pulsations have not been
detected (Karpova et al.~2015).

\section{Observations and Analysis}

We observed \src\ with {\sl Chandra} for 17.5 ks on 16 November 2014,
and for 32 ks on 28 November 2014 (obsIDs 16758 and 17551,
respectively), with the ACIS S3 CCD chip.  All data were reprocessed
with {\sl CIAO} v4.7 and CALDB v4.6.8, and screened for periods of
high particle background.  The absence of bright X-ray sources from
the ROSAT All-Sky Survey in the vicinity of \src\ ensured that any
emission would be of low surface brightness, so we used Very Faint
mode for more efficient background rejection.  Spectral analysis was
done with XSPEC v12.8.2 \citep{arnaud96}. Background was extracted
from a large area on the S3 CCD chip away from \src. The background
was modeled instead of subtracted in order to allow the use of Markov
chain Monte Carlo (MCMC) methods that is necessary for the unbiased
estimation of spectral model parameters for faint X-ray sources
\citep[e.g.,][]{vandyk01}. We assumed non-informative priors in
spectral fits, either uniform or logarithmic, with the latter used
only for (absorbed) fluxes.  No spectral binning was used, expect when
plotting spectra and model fits.

We detected a moderately bright source at the location ($\alpha$,
$\delta$) = ($3^h 37^m 12.86^s$, $61^\circ 53' 1.9''$), containing 241
counts, which we designate CXOU J033712.8 615302.  These coordinates
are the average of positions in individual pointings, $3^h 37^m
12.892^s$ ($3^h 37^m 12.824^s$) and $61^\circ 53' 2.01''$ ($61^\circ
53' 1.77''$) for observations 17551 (16758). They correspond to source
centroids estimated with help of the {\tt srcextent} tool in {\sl
  CIAO}. The source size estimated with {\tt srcextent} is consistent
with an unresolved point source.  The positional uncertainty is almost
entirely due to the {\sl Chandra} external astrometric errors
\citep[mean error of $0.''16$;][]{rots09}. Statistical errors are
significantly smaller, and equal to $0.''07$ ($0.''09$) for
observations 17551 (16758).  These errors were estimated from equation
(14) of \citet{kim07}. As there are no optical or radio counterparts
to this X-ray source (or to other X-ray sources sufficiently close to
the {\sl Chandra} optical axis to allow for reliable measurements of
their positions), a more accurate determination of the source position
is currently not possible.

The source location is near the peak brightness of the radio image
(Fig.~\ref{radio}).  Three other much fainter point sources can be
seen within the extent of the radio nebula.  Each has about 20 cts; if
they have the same spectral shape as \src, their fluxes are about $8
\times 10^{-15}$ erg cm$^{-2}$ s$^{-1}$ (see below).  At this level,
the number of sources per square degree found in the {\sl Chandra}
Deep Field South is about 400 \citep{lehmer12}, or about 0.1
arcmin$^{-2}$.  Since the radio extent of \src\ is over 8
arcmin$^{-2}$, we can be reasonably sure these are unassociated
background sources, probably AGNs which dominate the counts at this
flux level \citep{lehmer12}.

With this number of counts, only simple spectral fitting was possible.
We made MCMC fits with power-law and blackbody spectral distributions
in the energy range from 0.5 keV to 8 keV (all detected source counts
are within this energy range), allowing for absorption if
necessary. The power-law and blackbody fits are equally acceptable,
but the temperature of the blackbody ($0.93_{-0.08}^{+0.10}$ keV) is
unreasonably high for a putative neutron star with no evidence of
youth. Furthermore, no interstellar absorption is required for the
best-fit blackbody (95\%\ upper limit to $N_H$ is $2.0 \times 10^{21}$
cm$^{-2}$).  There is significant extinction
($E(B-V)=0.64_{-0.035}^{+0.033}$) within 1 kpc in this direction on
the sky \citep{green15}, so the blackbody fit is inconsistent with the
4 kpc distance derived from the \ion{H}{1} absorption measurements
toward \src.

\begin{figure}
  \includegraphics[scale=0.35]{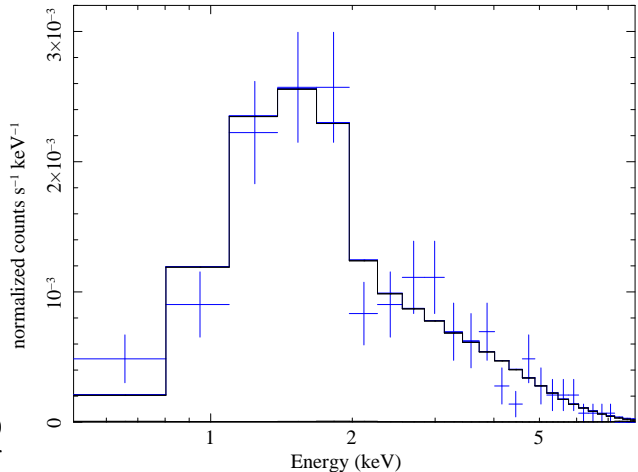}
  \caption{Spectrum of X-ray point source, with the best-fit power law
    model shown.}
  \label{fig_spectrum}
  \end{figure}

The power-law fit, shown in Figure~\ref{fig_spectrum}, gives an
absorbing column $N_H = 6.7 (4.0, 9.7) \times 10^{21}$ cm$^{-2}$
(using the Grevesse \& Sauval [1998] abundance set), reasonable if the
point source is at the 4 kpc distance of the radio PWN.  The power-law
index is $\Gamma = 1.8 (1.4, 2.2)$, quite typical for X-ray pulsars
(see, e.g., the catalog in Kargaltsev et al.~2013). The (absorbed)
flux between 0.5 and 8 keV is $6.1 (5.2, 7.1) \times 10^{-14}$ erg
cm$^{-2}$ s$^{-1}$.  After calculating unabsorbed fluxes for each MCMC
draw (i.e., a triple consisting of $N_H$, $\Gamma$, and the absorbed
flux), we arrived at an unabsorbed flux of $9.0 (7.6, 11.1) \times
10^{-14}$ erg cm$^{-2}$ s$^{-1}$, giving an unabsorbed luminosity
within this energy range of $L_X = 1.7 (1.4, 2.1) \times 10^{32}$ erg
s$^{-1}$ at 4 kpc.

The timing analysis was performed on observations that were corrected
to barycenter using the {\sl CIAO} task {\tt axbary} (with the source
coordinates listed above). We searched for pulsations in photon
arrival times in the frequency range from about $\nu_{min}=0.08$ Hz to
$\nu_{max}=0.159$ Hz. The minimum frequency corresponds to a rotation
period of 12.5 s. This is slower than the rotation period of all known
isolated neutron stars, including magnetars. The 3.141 s time
resolution of our observations sets the highest ($1/6.282\ {\rm s} =
0.159$ Hz) frequency to be searched. It also restricts our search to
purely sinusoidal signals.  We used the well-known Rayleigh ($Z_1^2$)
test instead of phase folding because of its higher sensitivity
\citep[e.~g., see][]{leahy83}. The number of independent frequency
searches is $3T(\nu_{max}-\nu_{min})$, where $T=301^h$ is the total
time elapsed between the beginning and end of {\sl Chandra}
observations of \src\ \citep[the factor of 3 accounts for oversampling
  in frequency for the $Z_1^2$-test;][]{dejager89}. The maximum
$Z_1^2$ power found, $Z_{1,{\rm max}}^2=19.8$, is not statistically
significant in view of the large number of frequencies searched.  This
$Z_{1,{\rm max}}^2$, in combination with the total number of counts in
the source extraction region, corresponds to an upper limit of 0.54
for the pulsed fraction \citep[at 95\%\ confidence; for this estimate,
  we used the method of][]{brazier94}.

There is no apparent extended emission near the point source.  In an
annulus with inner and outer radii of 2.5 pixels ($1.23\arcsec$) and
13 pixels ($6.4\arcsec$), respectively, there are 27 counts in the
0.5--8 keV energy range. We used the {\sl CIAO} task {\tt arfcorr} to
generate simple synthetic PSFs at several photon energies which were
then combined to make predictions for ratios of counts in this annulus
relative to an aperture with radius equal to the inner radius of the
annulus.  The measured counts within this aperture plus background
estimates far from the source served as input to a two-component model
consisting of a uniform background and a PSF model for the point
source. This model predicts 23 counts within this annulus, so there is
no evidence for diffuse emission from a PWN. Based on this model and
assuming the same spectral shape for the PWN and the point source, a
$3 \sigma$ upper limit to the (absorbed) 0.5-8 keV PWN surface
brightness near the point source is $3 \times 10^{-17}$ erg cm$^{-2}$
s$^{-1}$ arcsec$^{-2}$, or 6\%\ of the measured flux from the point
source when integrated over the annulus.

\section{Discussion}

Our discovery of the X-ray point source in \src\ essentially confirms
its identification as a PWN, in the rare class of steep-radio-spectrum
PWNe.  These objects also share properties of a point source (pulsar
in one case) near the geometric center of the radio nebula, and very
small X-ray nebulae (or, in the case of \src, only an upper limit).
The point sources in DA 495 and G76.9+1.0 differ considerably in
properties: in the former, the point source does not show pulsations
(pulsed fraction $\lapprox 40\%$; Karpova et al.~2015), but is well
fit by thermal models: a blackbody with $kT \sim 215$ eV, or a
magnetized neutron-star atmosphere model with $kT \sim 80 - 90$ eV.
The upper limit to a power-law contribution is $L_X \lapprox 5 \times
10^{31}$ erg s$^{-1}$.  In G76.9+1.0, the X-ray point source is
observed to be a pulsar, with a nonthermal spectrum with $\Gamma =
1.0$ and $L_X (2 - 10\ {\rm keV}) = 7 \times 10^{33}$ erg s$^{-1}$
\citep{arzoumian11}.  The point source in \src\ more closely resembles
the pulsar in G76.9+1.0, though much less luminous and with a slightly
steeper spectrum.

All evidence is consistent with our point source being the pulsar
powering the radio PWN.  (A search for radio pulsations has been
performed [D. Lorimer, PI] but nothing has been reported.)  The
adequacy of a power-law fit points to the emission we observe being
primarily magnetospheric in nature.  This suggests a short duty cycle
and high pulsed fraction for the pulsar (e.g., Kargaltsev \& Pavlov
2007). However, its properties are not unusual for X-ray
rotation-powered pulsars, and give no hint to the nature of the very
unusual radio PWN that surrounds it.

We see no trace of extended emission that could be a pulsar-wind
nebula.  For an assumed diameter of $13''$ (about 1/20 the radio
size), our $3 \sigma$ upper limit on the luminosity is about $10^{31}$
erg s$^{-1}$ (assuming the same spectrum as the pulsar), making any
PWN that size fainter than all but 5 of the 59 PWNe catalogued in
\cite{kargaltsev10}.  The ratio of X-ray to radio flux $S_X/S_r$ for
PWNe varies over a wide range, with a typical value being about 10
(e.g., Gaensler \& Slane 2006), but ranging to 100 or more (e.g., 600
for the PWN in G11.2--0.3; Tam et al.~2002, Kargaltsev et al.~2013).
A small early collection of radio-selected PWNe has $S_X/S_r$ ranging
from 10 to 1000 and above (Reynolds \& Chevalier 1984), while the
steep-spectrum PWN DA 495 has $S_X/S_r \sim 15$.  Roughly estimating
the integrated radio flux $S_r$ of G141.2+0.5 as $\nu S_\nu$ gives
$S_r \sim 2 \times 10^{-15}$ erg cm$^{-2}$ s$^{-1}$.  Then $S_X \sim
10 S_r$ would imply an X-ray flux of order $2 \times 10^{-14}$ erg
cm$^{-2}$ s$^{-1}$, about 1/5 of the flux in the point source, and
three times the upper limit we find for a presumed PWN diameter of
$26''$.  Alternatively, an extrapolation of the radio flux of 120 mJy
at 1.4 GHz to 1 keV ($2.4 \times 10^{17}$ Hz) with $\alpha = -0.69$
would imply a spectral flux of about 35 nJy there, or an integrated
flux between 0.5 and 8 keV (with the same spectral index) of $3 \times
10^{-13}$ erg cm$^{-2}$ s$^{-1}$, or about 20 times the flux in our
point source and far greater than any PWN flux.  We infer that, as
with all other known PWNe with radio to X-ray spectra, the spectrum
must steepen, even from its already steep radio value.  Whether this
is due to radiative losses or intrinsic spectral structure cannot be
determined at this time.

In both DA 495 and G76.9+1.0, the X-ray nebula is smaller than the
radio nebula by large factors (about 30 in each case).  The same ratio
in \src\ would mean an X-ray nebula with radius about $4''$.  We see
no evidence for extended emission beyond the PSF.  Any X-ray PWN could
be either smaller than 1/30 of the radio size, or considerably fainter
than 10 times the radio flux.  Either possibility would make \src\ a
unique pulsar/PWN combination in the Galaxy, whose further study might
provide important clues to the late-time PWN phenomenon.

This work was supported by NASA through the {\sl Chandra} General
Observer Program grant GO5-16055X. The scientific results reported in
this article are based on observations made by the {\sl Chandra} X-ray
Observatory.  This research has made use of software provided by the
{\sl Chandra} X-ray Center (CXC) in the applications packages {\sl
  CIAO} and {\sl ChIPS}.

\end{document}